\documentclass[12pt]{article}
\usepackage{graphicx}

\newcommand\pubnumber{CIPANP2015-Gollapinni}
\newcommand\pubdate{\today}

\def\ksu{Department of Physics\\
Kansas State University, Manhattan, Kansas, U.S.A.}
\def\support{\footnote{Work supported by U.S. DOE grant \textit{DE-SC0011840} and the Fermilab Intensity Frontier Fellowship.}}

\def\Title#1{\begin{center} {\Large #1 } \end{center}}
\def\Author#1{\begin{center}{ \sc #1} \end{center}}
\def\Address#1{\begin{center}{ \it #1} \end{center}}

\newcommand\pubblock{\rightline{\begin{tabular}{l} \pubnumber\\
         \pubdate  \end{tabular}}}
\newenvironment{Abstract}{\begin{quotation}  }{\end{quotation}}
\newenvironment{Presented}{\begin{quotation} \begin{center} 
             PRESENTED AT\end{center}\bigskip 
      \begin{center}\begin{large}}{\end{large}\end{center} \end{quotation}}
\def\Acknowledgements{\bigskip  \bigskip \begin{center} \begin{large}
             \bf ACKNOWLEDGEMENTS \end{large}\end{center}}

\def\beq{\begin{equation}}
\def\eeq#1{\label{#1}\end{equation}}
\def\eeqn{\end{equation}}

\def\beqa{\begin{eqnarray}}
\def\eeqa#1{\label{#1}\end{eqnarray}}
\def\eeqan{\end{eqnarray}}

\let\bar=\overbar

\def\Dslash{\not{\hbox{\kern-4pt $D$}}}
\def\dslash{\not{\hbox{\kern-2pt $\del$}}}

\def\msb{{\bar{\ssstyle M \kern -1pt S}}}

\begin{document}
\begin{titlepage}
\pubblock

\vfill
\Title{Accelerator-based Short-baseline Neutrino Oscillation Experiments}
\vfill
\Author{Sowjanya Gollapinni\support \\for the microboone collaboration}
\Address{\ksu}
\vfill
\begin{Abstract}
Over the last two decades, several experiments have reported anomalous results that could be hinting at the exciting possibility of  sterile neutrino states in the eV$^{2}$ mass scale. Liquid Argon Time Projection Chambers (LArTPCs) are a particularly promising technology to explore this physics due to their fine-grained tracking and exceptional calorimetric capabilities. The MicroBooNE experiment, a 170 ton LArTPC scheduled to start taking data very soon with Fermilab's Booster Neutrino Beam (BNB), will combine LArTPC development with the main physics goal of understanding the low-energy electromagnetic anomaly seen by the MiniBooNE experiment. Looking towards the future, MicroBooNE will become a part of the \textit{short-baseline neutrino} program which expands the physics capabilities of the BNB in many important ways by adding additional LArTPC detectors to search for light sterile neutrinos and bring a definitive resolution to the set of existing experimental anomalies. This paper will give an overview of the accelerator-based short-baseline neutrino oscillation program with a focus on the MicroBooNE experiment while highlighting the prospects for further addressing the short-baseline anomalies in the near future.
\end{Abstract}
\vfill
\begin{Presented}
Twelfth Conference on the Intersections of Particle and Nuclear Physics\\
Vail, Colorado,  May 19--24, 2015
\end{Presented}
\vfill
\end{titlepage}
\def\thefootnote{\fnsymbol{footnote}}
\setcounter{footnote}{0}

\section{Motivation}

The Standard Model (SM) of Particle Physics predicts three active neutrino flavors. This was later confirmed by the experiments on the Large Electron-Positron ring (LEP) ring that demonstrated that only three neutrino flavors couple to the Z boson~\cite{LEP}. As shown in Table \ref{tab:anamoly},  in the last two decades, several short-baseline ($<$1 km) oscillation experiments reported anomalous results that do not fit the SM 3-$\nu$ scenario.

As seen from the Table below, even though each measurement alone lacks discovery potential, together they could be hinting at new physics. The most common interpretation of these hints is high $\Delta m^{2}$ neutrino oscillations with one (or more) additional \textit{sterile}  
neutrino states with masses at or below a few eV$^{2}$. Sections 1.1 and 1.2 briefly discuss the anomalous results observed by the LSND and MiniBooNE experiments. 
\begin{table}[h]
\begin{center}
\begin{tabular}{|c|c|c|} 
\hline
Experiment  & Channel &  Significance \\ \hline 
LSND        &  Charged-current& 3.8$\sigma$ \\ 
&$\bar{\nu}_{\mu} \rightarrow \bar{\nu}_{e}$      &      \\ \hline
MiniBooNE   &  Charged-current     &     3.4$\sigma$      \\ 
& $\nu_{\mu} \rightarrow \nu_{e}$&\\ \hline
MiniBooNE        &  Charged-current& 2.8$\sigma$ \\ 
&$\nu_{\mu} \rightarrow \nu_{e}$      &      \\ \hline
GALLEX$/$SAGE &   $\nu_{e}$ disappearance    &    2.8$\sigma$  \\ \hline
Reactors   &   $\bar{\nu}_{e}$ disappearance   &     3.0$\sigma$ \\
\hline
\end{tabular}
\caption{Table listing interesting experimental anomalies from short-baseline neutrino and source experiments~\cite{sterile}. LSND  and MiniBooNE are accelerator-based short-baseline experiments. The rest are radioactive source and reactor experiments~\cite{reactor, kopp}.}
\label{tab:anamoly}
\end{center}
\end{table}
\subsection{High $\Delta m^{2}$ results: the LSND experiment}
The Liquid Scintillator Neutrino Detector (LSND) experiment~\cite{LSND} at Los Alamos lab was the first to observe evidence for neutrino oscillations beyond the SM 3-$\nu$ scenario. The LSND detector consisted of a cylindrical tank filled with 167 tons of mineral oil and 0.031 g/l of b-PBD organic scintillating material. An array of 1220 photomultiplier tubes covered the inner surface of the detector. The baseline (distance from the neutrino source to the detector) of the experiment was approximately 30~m.

\begin{figure}[htb]
\centering
\includegraphics[height=2.4in]{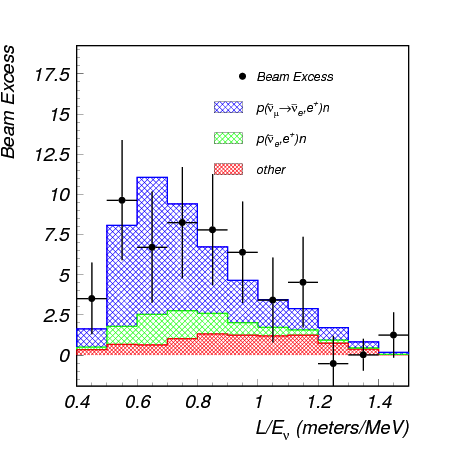}
\includegraphics[height=2.4in]{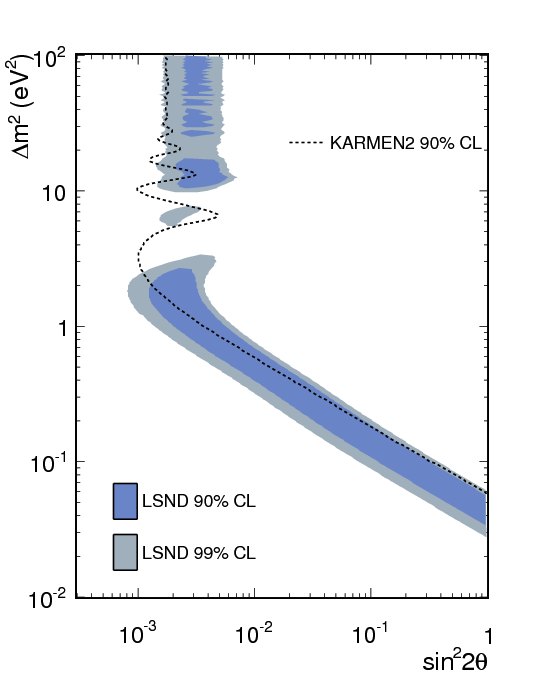}
\caption{ (Left) LSND event excess as a function of $L/E_{\nu}$ and comparison to the predicted oscillation signal. (Right) LSND allowed ($\Delta m^{2}$, $\sin^{2}2\theta$) parameter space and limits from KARMEN for a 2-$\nu$ oscillation model~\cite{excess,KARMEN}.}
\label{fig:lsnd}
\end{figure}

The LSND experiment primarily searched for $\bar{\nu}_{e}$ in a $\bar{\nu}_{\mu}$ beam originated by $\mu^{+}$ at rest with $E_{\nu}$ approximately between 0 and 53 MeV. The $\bar{\nu}_{e}$ candidate events are identified through the \textit{inverse beta decay} process ($\bar{\nu}_{e} p \rightarrow e^{+}n$) which resulted in a two-fold signature of a prompt positron emitting 52.8 MeV cherenkov radiation and a correlated 2.2 MeV $\gamma$ from neutron capture. The LSND experiment measured an excess of 87.9$\pm$23.2 $\bar{\nu}_{e}$ events (see Figure \ref{fig:lsnd} Left). Interpreting the excess as oscillations in a two-neutrino model (see Figure \ref{fig:lsnd} Right), LSND obtained 3.8$\sigma$ evidence for $\bar{\nu}_{\mu} \rightarrow \bar{\nu}_{e}$ oscillations with $\Delta m^{2} >$ 0.2 $eV^{2}$~\cite{excess}.

Worth mentioning here is the KARMEN experiment~\cite{KARMEN} (at a baseline of 17.7 m) that took data around the same time as the LSND experiment and saw no evidence of an oscillation signal. However, an analysis of LSND and KARMEN data~\cite{karmenlsnd} excluded the high $\Delta m^{2}$ ($>$ 10 $eV^{2}$) region and restricted the allowed region to $\Delta m^{2} < $ 1 $eV^{2}$ or $\Delta m^{2} \approx$ 7 $eV^{2}$ in the ($\Delta m^{2}$, $\sin^{2}{2\theta}$) space (see Figure \ref{fig:lsnd}).

\subsection{High $\Delta m^{2}$ results: the MiniBooNE experiment}
MiniBooNE, a 12m diameter Cherenkov detector~\cite{miniboone} filled with 800 tons of mineral oil was initially conceived as a test of the LSND anomaly. MiniBooNE sits on the Fermilab's Booster neutrino beam line  (BNB)~\cite{bnb} at a distance of 541~m from the neutrino source. In contrast to LSND, the $\nu_{\mu}$ ($\bar{\nu}_{\mu}$) beam from the Fermilab BNB is produced by $\pi^{+}$ and $K^{+}$ ($\pi^{-}$ and $K^{-}$) that decay-in-flight in a 50~m decay pipe and typically has an energy that peaks around 800 MeV. This results in a similar L/E ($\approx$ 1 m/MeV) as LSND but the signal, backgrounds and systematic uncertainties are very different than in LSND. Particles are identified in the MiniBooNE detector through their cherenkov light patterns. For example, particles such as electrons and photons produce a fuzzy Cherenkov ring, while muons produce a sharp outer ring with a fuzzy inner region, neutral pions that make two electron/photon like tracks result in two fuzzy rings and so on. The $\nu_{e}$-like candidate events are identified by a single isolated electron-like cherenkov ring.

\begin{figure}[htb]
\centering
\includegraphics[height=2.5in]{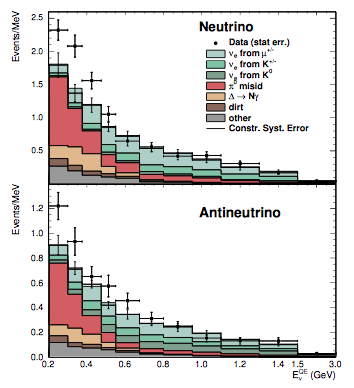}
\includegraphics[height=2.5in]{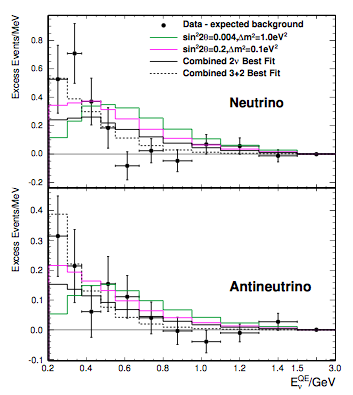}
\includegraphics[height=2.5in]{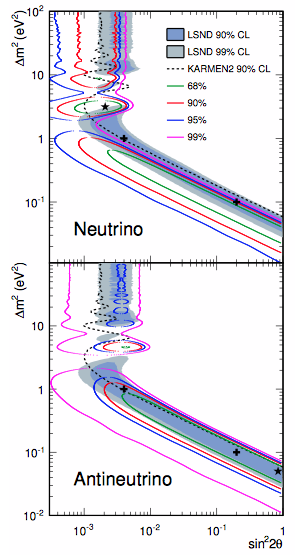}
\caption{ (Left) The neutrino mode (top) and antineutrino mode (bottom) reconstructed neutrino energy distribution for $\nu_{e}$ CCQE candidate events. (Center) Comparison of CCQE event excesses 
in neutrino (top) and antineutrino mode (bottom) to the 2-$\nu$ and $3+2$ oscillation best fits. Also shown are the two reference values (magenta and green lines) in the LSND allowed region.  (Right) MiniBooNE  combined neutrino and anti-neutrino mode allowed regions in ($\Delta m^{2}$, $sin^{2}2\theta$) phase space for candidate CCQE events with energies between 200 and 3000 MeV. The black star shows the best fit point for a 2-$\nu$ osicllation model~\cite{minibooneexcess}.}
\label{fig:minib}
\end{figure}

The MiniBooNE experiment ran for 10 years (2002 to 2012) switching between neutrino and anti-neutrino modes. The first MiniBooNE data taking started in neutrino mode since it was understood that the anti-neutrino mode suffers from higher backgrounds especially coming from $\nu_{\mu}$ contamination and running in the neutrino mode prior to anti-neutrino mode gave the opportunity to understand and constrain backgrounds directly from measurements in the detector~\cite{ccqe, antinumu} and achieve higher statistics. The final MiniBooNE oscillation results~\cite{minibooneexcess} showed an excess in both $\nu$ (3.4$\sigma$) and $\bar{\nu}$ (2.8$\sigma$) modes in the low energy region (below 475 MeV). The excess in neutrino mode is only marginally compatible with a simple 2-$\nu$ oscillation hypothesis while the excess in the anti-neutrino mode is consistent with anti-$\nu$ oscillations in the $0.01<\Delta m^{2}<1.0$ $eV^{2}$ range with some overlap with the LSND signal (see Figure \ref{fig:minib}).

Although MiniBooNE greatly constrained various backgrounds by direct measurements and observed a statistically significant excess, there are some limiting factors to the MiniBooNE oscillation analysis. The MiniBooNE detector being a Cherenkov detector cannot distinguish electrons (signal events) from single photons (background events). The measured signal in neutrino mode is only marginally compatible with the expected sterile neutrino signal. Any further test of the MiniBooNE signal requires a highly granular detector that can distinguish between photons and electrons and obtain more information on the neutrino interactions themselves.

\begin{figure}[t]
\centering
\includegraphics[height=3.0in, width=3.5in]{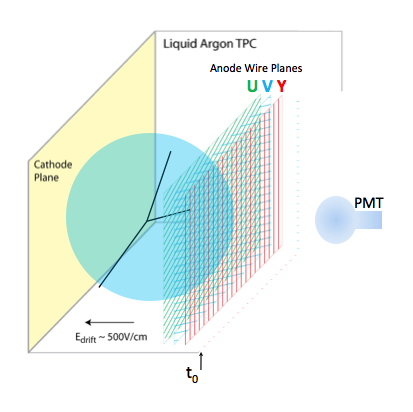}
\caption{Basic principle of a LArTPC.}
\label{fig:lartpc}
\end{figure}

\section{Liquid Argon Time Projection Chambers}
Liquid Argon Time Projection Chambers (LArTPCs) are imaging detectors that offer exceptional calorimetric and position resolution capabilities for studying neutrino interactions in argon and are rapidly evolving as a desirable detector technology for future neutrino experiments. The idea to use LArTPC technique for particle detection was first proposed in the 1970s~\cite{radeka, crubbia} and a tremendous amount of progress has been made since then in terms of developing this technology~\cite{chen,icarus,argoneut}. A LArTPC typically consists of a cathode plane and finely segmented anode planes enclosed in a volume of highly purified liquid argon (LAr) (see Figure \ref{fig:lartpc}). Neutrino interactions with LAr in the TPC produces charged particles that cause the ionization and excitation of the argon. A large electric field drifts electrons towards finely segmented (mm-scale) anode wire planes oriented at different angles to provide stereoscopic views of the same interaction. The excitation of argon produces prompt scintillation light giving important timing information about the neutrino interaction. 

\subsection{$e^{-}/\gamma$ separation in a LArTPC}
LArTPCs combine topology and energy deposition information along the track (dE/dx) to distinguish electrons from photons.
\begin{itemize}
\item In the case of a $\gamma$ or $\pi^{0}$, since they are neutral, one can look for a gap between the vertex and electro-magnetic shower
\item If no gap is observed, one can measure the charge deposition (dE/dx) at the start of the shower (first few cm of the shower). If the measured dE/dx corresponds to 2.1 MeV/cm, equivalent to a minimum ionizing particle (MIP) in argon, then it is more likely an $e$-like candidate. If the measured dE/dx corresponds to twice the MIP deposition (4.2 MeV/cm), then it is $\gamma$-like since a $\gamma$ pair produces.
\end{itemize}

\begin{figure}[htb]
\centering
\includegraphics[height=2.3in]{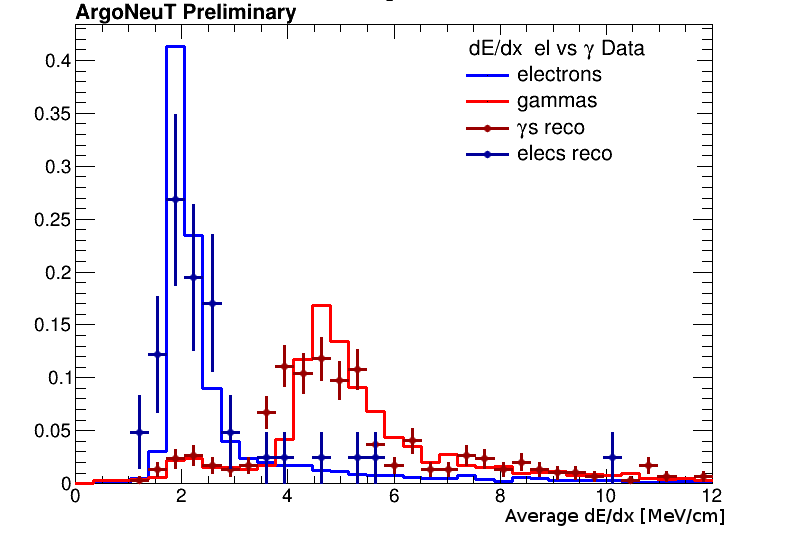}
\caption{$e^{-}/\gamma$ separation measured in ArgoNeuT~\cite{dedxproof}. The solid blue/red lines represent MC based separation and the blue/red points represent separation based on topology and measured dE/dx.}
\label{fig:argoneut}
\end{figure}

\noindent The ArgoNeuT detector~\cite{argoneut} on the NuMI (Neutrinos from the Main Injector) beamline at Fermilab served as the proof of principle for the above strategy as shown in Figure \ref{fig:argoneut}~\cite{dedxproof}. 

\subsection{The MicroBooNE LArTPC}
MicroBooNE is a new 170-ton LArTPC neutrino experiment (largest so far in the U.S.) built on the Fermilab Booster neutrino beamline. MicroBooNE is an important step towards LArTPC R$\&$D in establishing large-scale detectors for neutrino physics. MicroBooNE is currently commissioning and will start taking neutrino data very soon. MicroBooNE brings several technological advances including argon purification without evacuation, cold (in argon) front-end electronics, and a large (2.5~m) electron drift.
Physics-wise, MicroBooNE has two main goals: 1). Investigate the MiniBooNE low energy excess and 2). Produce first high-statistics precision measurements of $\nu$-Ar interactions in the 1 GeV range. The MicroBooNE detector sits at a baseline of 470~m on the BNB yielding an L/E of approximately 1 m/MeV similar to that of MiniBooNE.

The MicroBooNE LArTPC consists of approximately 80 tons (active volume) of liquid argon. The distance between cathode and anode is 2.56~m and an ionization electron takes about 1.6~ms to travel the full drift distance. The anode region consists of 3 wire planes oriented at different angles with a total of 8256 wires (each wire is 150 $\mu$m thick). The spacing between consecutive wires and wire planes is 3~mm. The MicroBooNE light collection system consists of 32 8-inch PMTs that are located just behind the wire planes and detect scintillation light from $\nu$-Ar interactions. The PMT information is used to trigger on beam events and significantly reduce the data throughput.

It is important to note here that in the 1 GeV energy range, several neutrino processes contribute and nuclear effects are large. Understanding $\nu$-Ar cross-sections over the energy range valid for short and long baseline experiments is vital for any oscillation measurement and MicroBooNE will provide valuable information on these low energy $\nu$-Ar cross-sections.

Given the fine granularity and exceptional calorimetric capabilities, MicroBooNE will be capable of telling whether the MiniBooNE excess is electron-like or photon-like. If the excess is electron-like, it will likely indicate beyond the SM oscillations involving sterile neutrinos. If the excess is photon-like, it will likely indicate a new unmodeled source of background impacting $\nu_{e}$ appearance experiments. likely indicate new unknown cross-sections. While MicroBooNE can address a critical piece of the short-baseline puzzle, MicroBooNE by itself is not large enough to explore the complete sterile neutrino oscillation parameter space. This motivated the idea of a combined short-baseline neutrino (SBN) program at Fermilab.

\section{Short-baseline Neutrino Program at Fermilab}
The SBN program at Fermilab builds upon the already existing MicroBooNE detector by adding additional detectors along the BNB. 
The Short-baseline Near Detector (SBND) with 112 tons of active LAr mass  will sit closest to the Booster Neutrino source at 110~m. The 
main physics goal of SBND is to characterize the intrinsic BNB content before any oscillations occur. The SBND construction is expected to start soon. The ICARUS-T600 detector with 476 tons of active LAr mass will move from CERN to Fermilab in 2017 and will act as the far detector sitting at 600~m on the BNB. Together with OPERA experiment, ICARUS already made important contributions to the sterile neutrino search by limiting the window of LSND anomaly. 

The two main advantages of using the Booster neutrino source for the SBN program are the following: 1). BNB produces a shallow beam, approximately 10~m detector hall depth at all baselines (no additional costs incurred for specialized underground buildings but the surface location results in more cosmic background) and 2). a well understood beam (more than 10+ years of experience from MiniBooNE and SciBooNE, also from HARP and BNL E910 experiments). 
This multiple-detector arrangement allows a definitive search for sterile neutrinos in the region where there are existing hints. Using the same neutrino source (BNB) and requiring that all three detectors have the same detector technology will significantly reduce systematic uncertainties to \% level.

\begin{figure}[b!]
\centering
\includegraphics[height=2.3in]{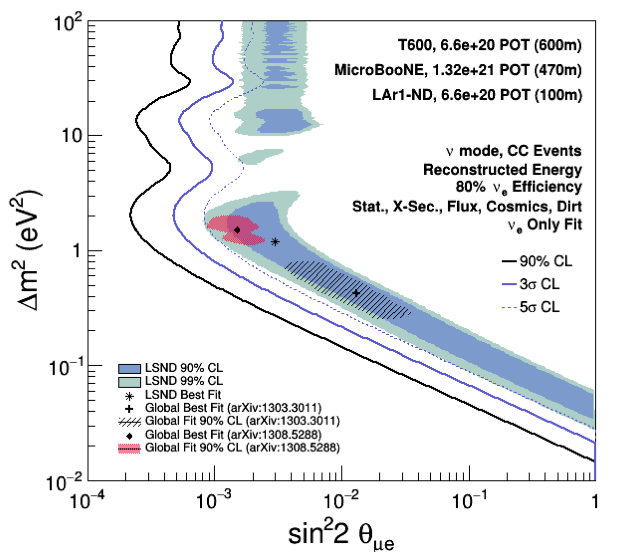}
\includegraphics[height=2.3in]{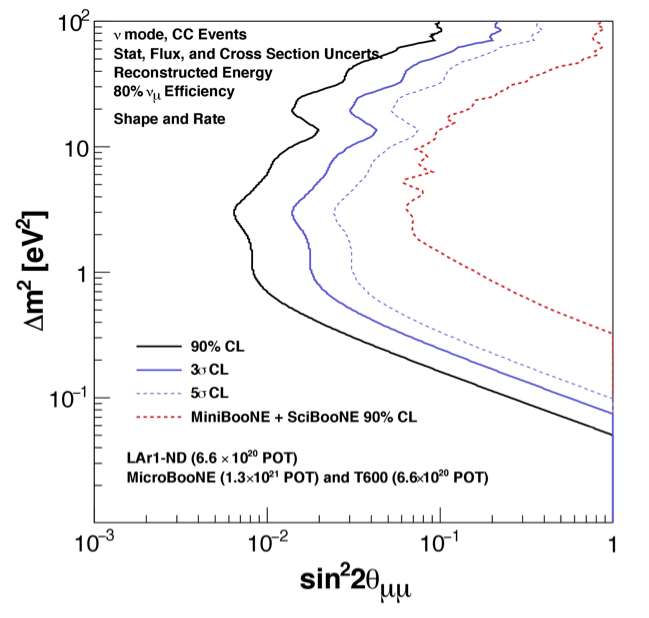}
\caption{Sensitivity of the SBN Program to $\nu_{\mu} \rightarrow \nu_{e}$ appearance~\cite{sbn} (left). Sensitivity of the SBN Program to $\nu_{\mu} \rightarrow \nu_{x}$ disappearance~\cite{sbn} (right). In both cases, sensitivity analysis is done assuming a $6.6\times10^{20}$ protons on target (POT) exposure in SBND and ICARUS-T600 and $13.2\times10^{20}$ POT in MicroBooNE (since MicroBooNE will have commenced its data-taking earlier).}
\label{fig:sensi}
\end{figure}

\subsection{SBN Oscillation searches}
Global analysis of short-baseline neutrino results from Giunti \textit{et al.}~\cite{giunti} and Kopp \textit {et al.}~\cite{kopp} show that the allowed parameter regions for
neutrino and anti-neutrino data indicate preferred $\Delta m^{2}_{41}$ values in the ∼[0.2$-$2] $eV^{2}$ range. Multiple LArTPC detectors at different baselines along the BNB (as discussed in the previous section) will allow a very sensitive search of neutrino oscillations in multiple channels in the current experimental $\Delta m^{2}$ landscape. The sterile neutrino search through $\nu_{\mu} \rightarrow \nu_{e}$ appearance and $\nu_{\mu} \rightarrow \nu_{x}$ disappearance constitute the flagship measurements of the SBN program. A detailed Monte Carlo study was performed in order to understand the combined sensitivity of the SBN experiments. This study is explained in detail in Ref. 21.

Figure \ref{fig:sensi} (left) shows the sensitivity of the SBN Program to $\nu_{\mu} \rightarrow \nu_{e}$ appearance oscillation signals~\cite{sbn}. One can see from the figure that the LSND 99\% C.L. allowed region is covered at $\geq$ 5$\sigma$ level above $\Delta m^{2} = 0.1$ $eV^{2}$ (please note that the region below $\Delta m^{2} = 0.1$ $eV^{2}$ is already ruled out~\cite{kopp, giunti}). A $\nu_{e}$ appearance signal should be accompanied by a corresponding $\nu_{\mu}$ disappearance signal with equal or higher probability  in order to confirm any $\nu_{e}$ appearance oscillation signal that will be observed. Figure \ref{fig:sensi} (right) shows the sensitivity of the SBN Program to $\nu_{\mu} \rightarrow \nu_{x}$ disappearance~\cite{sbn}. Comparing the red curve and the solid black curve in the Figure, one can see that the SBN program can extend the search for $\nu_{\mu}$ disappearance an order of magnitude beyond the combined analysis of SciBooNE and MiniBooNE.

\section{Summary}
The SBN program at Fermilab is well positioned to explore the sterile neutrino parameter space where there are existing hints. Existence of sterile neutrinos would be a revolutionary discovery. The first experiment in the SBN program, MicroBooNE, will start taking data very soon and in addition to addressing the MiniBooNE low-energy excess, MicroBooNE will advance the powerful LArTPC technology for future multi-kiloton detectors and also perform precision $\nu$-Ar cross-section measurements. 

\Acknowledgements
The author thanks the organizers of CIPANP 2015 conference for the invitation to the conference.

\end{document}